\documentclass[11pt]{ar}
\usepackage[english]{babel}
\usepackage{psfig,graphicx}

\onecolumn

\begin{document}

\title{Kinematic structure of the atmosphere and envelope of the post-AGB star
      HD\,56126}

\author{Valentina Klochkova \& Eugenij Chentsov}

\institute{Special Astrophysical Observatory RAS, Nizhnij Arkhyz,  Russia} 

\date{\today}	     

\abstract{We present results of an analysis of the optical spectrum of the
post-AGB star HD\,56126 (IRAS\,07134\,+\,1005) based on observations made
with the echelle spectrographs of the 6-m telescope with spectral
resolutions of R\,=\,25000 and 60000 at 4012--8790\,\AA. The profiles of
strongest lines (HI; FeII, YII, BaII absorptions, etc.) formed in the
expanding atmosphere at the base of the stellar wind have complex and
variable shapes. To study the kinematics of the atmosphere, the velocities
of individual features in these profiles must be measured. Differential
line shifts of up to $\rm V_r$\,=\,15$\div$30\,km/s have been detected
from the lines of metals and molecular features. The star's atmosphere
simultaneously contains both expanding layers and layers falling onto the
star. A comparison of the data for different times demonstrates that both
the radial velocity and the velocity pattern in whole are variable. The
position of the molecular spectrum is stable, implying stability of the
expansion velocity of the circumstellar envelope around HD\,56126 detected
in observations in the C$_2$ and NaI lines.}

\authorrunning{Klochkova \& Chentsov}
\titlerunning{HD\,56126: structure of the atmosphere and envelope}

\maketitle

\section{Introduction}

We studied the kinematic parameters of the atmosphere of the star
HD\,56126 (SAO\,96709), which belongs to the asymptotic giant branch (the
post-AGB stage). In this short evolution stage, stars are in the process
of their transition from the AGB to becoming a planetary nebula; for this
reason, they are generally known as proto-planetary nebulae (PPNe). In the
Hertzsprung--Russell diagram, post-AGB stars evolve toward the left from
the AGB, maintaining nearly constant luminosity while becoming hotter. As
descendants of AGB stars, these objects can be used to trace variations in
the physical conditions and chemical parameters of the stellar material
due to changes in the sources of energy release in the stellar interior,
the ejection of the envelope, and mixing.

HD\,56126 is the optical component of the IR source IRAS\,07134+1005,
which has a double peaked spectral energy distribution (SED), typical for
PPNe. In addition to this anomalous SED, associated with the circumstellar
dust envelope, the star also displays other characteristics of this type
of object [\cite{Kwok}]: the optical component of the PPN is a F5\,Iab
supergiant outside the galactic plane (b=+9$\lefteqn{.}^m$99); the central
star is surrounded by an extended nebula whose angular size exceeds
4$^{\rm ''}$, according to Hubble Space Telescope observations
[\cite{Ueta}] (the largest known for this type of PPN); and the optical
spectrum displays H$\alpha$ emission and absorption with a variable line
profile [\cite{Oud1994}]. In addition to these general PPN characteristics
noted by Kwok [\cite{Kwok}], subsequent studies of HD\,56126 and the
associated IR source revealed several important peculiarities expected for
this evolution stage: a large excess of carbon and $s$-process elements
[\cite{Kloch1995}] and an emission feature at $\lambda$\,=\,21\,$\mu$ in
the IR--spectrum.

In the small subgroup of PPN having this emission, the presence of this
21\,$\mu$ feature was found to correlate with observational manifestations
of the products of nucleosynthesis (the excess of carbon and heavy metals
in the outer layers of the atmosphere) [\cite{Kloch1998, Decin}]. Thus,
HD\,56126 displays all the properties ex pected for a post-AGB object,
indicating the importance of detailed studies of its optical spectrum with
high spectral resolution in a broad wavelength range. Fortunately,
HD\,56126 is fairly bright (B\,=\,9$\lefteqn{.}^m$11, V = 8.27m), making
it the most convenient carbon enriched PPN star for high resolution
spectroscopy.

The main moments of our study are spectral features identification and
comparison the spectra of HD\,56126 and the standard $\alpha$\,Per
(Sp\,=\,F5\,Iab); search for profile variability for spectral features;
analyses the radial velocities $\rm V_r$ to search for differential
shifts; and studying the variability of the radial velocity. In Section~2,
we briefly describe the techniques used for the observations, processing,
and analysis of the spectral data. The main conclusions concerning the
spectrum of the star are presented in Section~3.1, while Sections~3.2 and
3.3 present our radial velocity measurements derived using various
features of the spectrum, and discuss the temporal behavior of the
radial-velocity pattern. Our results are summarized in Section~4.

{\footnotesize
\begin{table}[t]
\caption{Moments of observations and values of heliocentric radial velocitity
        $\rm V_r$ measured.
        Column~4 contains $\rm V_r$  values averaged over weak lines
        (with depths R$_\lambda$ close to the continuum level, 
	R$_\lambda \rightarrow$0).
        Velocities corresponding to the positions of the strongest
        components are presented for FeII(42), H$\alpha$, and D2\,NaI,
        with values determined from the weakest components given in parantheses.
        The two velocities in italics in column~5 are determined from the IR--oxygen
        triplet OI\,7773\,\AA. Uncertain values are marked by a colon.}
\medskip                                              
\begin{tabular}{ll c| c  c  c  l  l l @{\quad} | l l l}                 
\hline
\small Date &\small  Spectro-- &\small  $\Delta\lambda$  & \multicolumn{7}{c}{\small  $V_r$} \\
\cline{4-12}
&\small graph  &\small  \AA{} &  R$_\lambda \rightarrow $0 & \small  Fe{\sc ii} &\small  H$\beta$ &
              \quad \small  H$\alpha$ & \small  D\,Na{\sc i}& \small C$_2$ &\multicolumn{3}{c}{\small  intestellar} \\
\hline
\quad  1& 2& 3&4&5&6&\quad7& 8&9& 10& 11 & 12\\
\hline
12.01.93&Lynx&5560--8790& 88.8 &{\it 91} &    &78 (100:)&77        &79: &     &     &    \\
10.03.93&Lynx&5560--8790& 89.0 &{\it 93} &    &71 (43:) &75:       &76: &     &     &     \\
04.03.99&Lynx&5050--6640& 85.9 & 77      &    &76 (43:) &78        &77.1&     &     &     \\
20.11.02&NES &4560--5995& 89.6 &95 (80:) &89  &         &74.9 (89) &77.2&12.0 &23.5 & 30.8\\
21.02.03&NES &5150--6660& 88.8 & 96:     &    &88 (112:)&75.6 (89) &77.1&12   &24   & 31 \\
12.04.03&NES &5270--6760& 88.4 &         &    &82 (103:)&75.4 (89:)&    &13   &23   & 30.5 \\
14.11.03&NES &4518--6000& 85.3 &96 (86:) &97  &         &75.0 (87:)&76.9&12.5 &     &       \\
10.01.04&NES &5270--6760& 86.7 &         &    &54: (78:)&75.6 (86:)&    &13.0 &23.5 & 31  \\
09.03.04&NES &5275--6767& 89.8 &         &    &58 (74:) &76.1 (89) &    &13   &24   & 31   \\
12.11.05&NES &4010--5460& 82.5 &97 (77:) &98  &         &          &77.5&     &     &      \\
\hline
\end{tabular}
%\label{spectra}
\end{table}
}

\section{Observations and reduction of the spectra}

We observed HD\,56126 and $\alpha$\,Per with the 6-meter telescope of the
Special Astrophysical Observatory. All the spectra were obtained at the
Nasmyth focus with the NES [\cite{Panch1999, Panch2002}] and Lynx
[\cite{Prep137, Prep139}] echelle spectrographs. In combination with a
2048$\times$2048 CCD and the image cutter [\cite{slicer}], NES provides a
spectral resolution of R\,=\,60000. The Lynx spectrograph used with a
1K$\times$1K CCD yields R\,=\,25000. The Table presents the observing
dates and wavelength regions detected. The spectral data were extracted
from two-dimension echelle-spectra using the ECHELLE context in the MIDAS
package modified to take into account specific features of the echelle
frames obtained with these spectrographs (see [\cite{Yushkin}] for
details). Cosmic-ray traces were removed via median averaging of pairs of
consecutive spectra. The wavelength calibtration was carried out using
spectra of a hollow cathode Th-Ar lamp.

The further processing, including photometric and position measurements,
was performed using the DECH20 code [\cite{gala}], which makes it possible
to find the positions of spectral features by overlapping their direct and
mirrored profiles. To increase the accuracy of the radial-velocity
measurement, the least blended lines in the observed and corresponding
synthetic spectra were compared.

In each spectrogram, the positional zero point was determined in the
standard way, via calibration using the positions of ionospheric night-sky
emission lines and telluric absorptions in the star spectrum. The accuracy
of the velocity measured from {\bf one line} in the spectra obtained with
the NES and Lynx spectrographs is about 1.0 and 1.5\,km/s, respectively.

\section{Results}

A list of the lines we identified in the spectrum of HD\,56126 is presented
at http://www.sao.ru/hq/ssl/HD56126-Atlas/Atlas.html. This contains only
a partial set of the lines for which the depths R and heliocentric radial
velocities $\rm V_r$ were measured. Let us consider in more detail the
peculiarities of the spectrum.

\subsection{Peculiarities of the spectrum of HD\,56126}

{\it Interstellar features.} Comparing our data with those obtained by
other authors, we tested the coincidence of the radial-velocity zero
points using interstellar and circumstellar lines. Figure~1 presents the
D2 NaI line profiles; both here and in the Table, to display the fine
structure of the lines, we consider only spectra obtained with our maximum
resolution, R\,=\,60000. As a result, five components of the D2 NaI line can
be distinguished. Three blue-shifted interstellar components of the D2
line in the spectrum of HD\,56126, clearly seen in Fig.\,1, yield the same
$\rm V_r$ values within the errors (columns~10--12 in the Table, with
the averaged values $\rm V_r$\,=\,12.6, 23.6, and 30.9\,km/s). The stability
of these values confirms that these three components are formed in the
interstellar medium. The shift of the component of the D2 NaI line with
$\rm V_r$\,=\,75.8\,km/s is consistent with that of the Swan bands of the C$_2$
molecule (columns (8) and (9) of the Table), which indicates that this
component is formed in the circumstellar medium. It is obvious that the
longest-wavelength component, $\rm V_r$\,=\,88.5\,km/s, is photospheric: its
temporal behavior corresponds to that of other photospheric absorption
features. Note that such a structure of the D NaI lines is
consistent with the results obtained by Bakker et al. [\cite{Bakker}], who
distinguished a weak component of the D2 NaI line with
$\rm <V_r>$\,=\,44\,km/s, which is also seen in at least three of our spectra,
yielding $\rm <V_r>$\,=\,46$\pm 1$\,km/s.
In addition, as we can see from Fig.\,1, the blend of three blue-shifted
components displays sharp boundaries, making it possible to reliably measure
the velocity for the blend as a whole. The average value derived from our data,
$\rm V_r$\,=\,20.3$\pm$0.3\,km/s, coincides with that obtained
by L'ebre et al. [\cite{Lebre}] from spectra with
lower resolution: $\rm V_r$\,=\,20$\pm$2\,km/s.

As was shown by Crawford and Barlow [\cite{Crawford}], when the C$_2$ and
KI circumstellar features are observed with ultra-high resolution, they
split into components spaced at about 1\,km/s. These components yield the
same set of velocities, but display different relative intensities. This
may result in small systematic differences (also of the order of 1\,km/s)
in the velocities obtained for the atomic and molecular circumstellar
lines at lower resolution. Our measurements do not indicate variations in
these velocities, while the average values 77.2$\pm$0.5\,km/s for C$_2$
and 75.4$\pm$0.3\,km/s for NaI do not disagree systematically with those
obtained by L'ebre et al. [\cite{Lebre}], Bakker et al. [\cite{Bakker,
Bakker1997}], and Crawford and Barlow [\cite{Crawford}]:
77.3$\div$77.6\,km/s and 75.3$\div$76.8\,km/s for C$_2$ and NaI, KI,
respectively.

{\it H$\alpha$ line in the spectrum of HD\,56126.} The H$\alpha$ lines in
the spectra of typical PPNe display complex (emission + absorption)
variable profiles with asymmetrical core, P~Cygni type profiles of various
kinds from direct to inverted, profiles with asymmetrical emission wings,
or profiles containing two emission components. A combination of such
features is also frequently observed. As we can see from Fig.\,2, the
H$\alpha$ line in the spectrum of HD\,56126 has both absorption and
emission components, which are not seen in the spectrum of the normal
supergiant $\alpha$\,Per. Figure~2 also clearly shows the H$\alpha$
absorption wings, extending to values close to those for the H$\alpha$
wings in the spectrum of $\alpha$\,Per. Figure~3, which presents only the
central parts of the H$\alpha$ profiles for all our observations,
demonstrates the profile variability from night to night.

Previously, based on their spectral monitoring of HD\,56126, Oudmaijer and
Bakker [\cite{Oud1994}] concluded that the H$\alpha$ profile is strongly
variable on timescales of two months. The analysis of the variability of
the H$\alpha$ profile and the corresponding set of radial velocities using
Fourier methods led L'ebre et al. [\cite{Lebre}] to conclude that there are
complex, pulsation-driven, dynamical conditions in the atmosphere of
HD\,56126. The extensive set of high-quality spectral observations of
HD\,56126 over almost eight years of Barth'es et al. [\cite{Barthes}]
indicated that, apart from H$\alpha$, the H$\beta$ line also varies.
Analyzing the profile variability of both these hydrogen lines, these
authors concluded that the observed variability does not display any
periodicity that could be related to variations of the radial velocity and
brightness of the star.

{\it Swan bands.} In addition to specific features of the HI line
profiles, the peculiarity of PPNe optical spectra is also manifest in the
presence of numerous molecular absorption features, along with lines
characteristic of F--K supergiants. In the spectrum of HD\,56126, with an
effective temperature Teff\,=\,7000\,K [\cite{Kloch1995}],
the C$_2$ Swan absorption band system is observed, as well as the red CN
system, first identified by Bakker et al. [\cite{Bakker}].
Later, molecular bands in the spectra of PPNe in another sample
(including HD\,56126), selected for the presence of the carbon-containing
molecules C$_2$, CN, and CH$^+$ in their envelopes, were studied in detail by
Bakker et al. [\cite{Bakker1997}] using high-resolution spectra (R\,=\,50000).
Judging from the velocity corresponding to the position of these bands, the
absorption molecular spectrum is formed in a restricted region of the
envelope, close to the star [\cite{Bakker1997}].

Our spectra of HD\,56126 contain several Swan bands. Figure~4 shows a
spectral fragment with the head of the C$_2$~(0;0) band at 5165\,\AA{}, and
compares again the spectra of HD\,56126 and the normal supergiant
$\alpha$\,Per. We present a detailed list of Swan band spectral features in
the spectrum of HD\,56126 and their intensities and corresponding radial
velocities at http://www.sao.ru/hq/ssl/HD56126-Atlas/ Atlas.html. The
radial velocities derived from the Swan bands are analyzed in Section\,3.2.
Although Swan bands have been detected in emission in the spectra of
several PPNe [\cite{Kloch1998, Bakker1997}], no signs of emission in
these bands or in the D lines of NaI have been found in spectra of
HD\,56126 obtained in various years. This is consistent with the fairly
simple elliptical shape of the nebula surrounding HD\,56126.

Apparently, emission in the Swan bands or the D lines of NaI is observed
only in the spectra of PPNe with bright circumstellar nebulae displaying
pronounced asymmetry. This is confirmed by the spectroscopic results for
IRAS\,04296+3429 [\cite{Kloch1999}], IRAS\,23304+6147 [\cite{Kloch2000}],
AFGL\,2688 [\cite{Kloch2000b}], IRAS\,08005$-$2356 [\cite{Kloch2004}],
IRAS\,20056+1834 [\cite{Rao, QYSge}], and IRAS\,20508+2011
[\cite{Kloch2006}]. As a rule, the HST images of these PPNe [\cite{Ueta}]
display bipolar structure. We emphasize that most of these objects belong
to Type~1 (PPNe with polarized optical radiation) according to the
classification suggested by Trammell et al. [\cite{Tram}]. 

Based on the parameters of the PPNe whose spectra we have studied,
including HD\,56126, we suggest that emission in the Swan bands and/or NaI
D lines is observed in the spectra of PPNe (and related stars with
envelopes) in which the outward propagation of the radiation of the central
star is substantially hindered by absorption in the circumstellar dust
envelope. Profiles of metal lines. Figures~5 and 6 show that the profiles
of strong lines of metals in the spectrum of HD\,56126 are not
symmetrical. In addition, comparisons of spectra obtained on different
nights indicate that, for the metallic lines formed in the expanding
atmosphere of the star (at the base of the wind), the profile shape varies
with both time and line intensity. Parthasarathy et al. [\cite{Parth}] had
already noted the asymmetry of the BaII\,6141\,\AA{} line profile in the
HD\,56126 spectrum. It cannot be ruled out that the variability of the
profiles is due to splitting of the absorption lines into components whose
positions and intensities are different at different times.
High-spectral-resolution observations reveal the multicomponent structure
of metal lines in the spectra of various types of variable stars, such as
classical Cepheids (see, for example, [\cite{Mathias}]) or pulsating
RV\,Tau [\cite{Gillet}] or W\,Vir type [\cite{Waller}] stars, due to the
presence of shocks (see also references below). The possibility for
several shocks to coexist in the atmosphere of HD\,56126 was shown by
Barth'es et al. [\cite{Barthes}]. However, direct observation of the
splitting of metal lines in the spectrum of HD\,56126 is complicated by
broadening in the turbulent atmosphere of the supergiant.

\subsection{Variability of the radial velocity pattern}

Some of PPNe display variability of the radial velocity $\rm V_r$ with
characteristic timescales of several hundred days, possibly providing
evidence for binarity. Indeed, conclusive evidence for orbital motion has
been obtained for several optically bright objects with IR--excesses. For
example, the binarity of the high-lattitude supergiants 89\,Her
[\cite{Ferro, Waters}] and HR\,4049 [\cite{Waelk}] has been proven, the
corresponding orbit elements determined, and models of the systems
proposed. Van Winckel et al. [\cite{Winckel}] showed that HR\,4049,
HD\,44179, and HD\,52961 are spectral binaries with orbital periods of
about one to two years. Bakker et al. [\cite{Bakker1998}] studied the
variability of complex emission--absorption profiles of the NaI D lines
and H$\alpha$ for HR\,4049 as functions of the orbital period using high
resolution spectra. The individual components of these lines are formed
either in the atmosphere of the primary, in the disk in which both binary
components are submerged, or in the interstellar medium. The nature of the
companions in suspected post-AGB binaries is still unknown, since there
are no directly manifestations of their radiation in either the continuum
or in spectral lines: all known post-AGB binaries are of type SB1. The
companions could be very hot or very low luminosity main sequence objects;
white dwarfs cannot be ruled out either, as in the case of BaII stars. The
$\rm V_r$ variability in PPNe can also be complicated by differential
motions in their extended atmospheres. A detailed analysis of $\rm V_r$
for selected bright PPNe using spectra with high spectral and temporal
resolution reveals differences in the behavior of $\rm V_r$ determined
from lines with different degrees of excitation, formed at different
depths in the stellar atmosphere.

In the following Section, we point out such differences in the behavior of
$\rm V_r$ for HD\,56126. Variability of the radial velocity of HD\,56126
was first suspected in [\cite{Kloch1995}], based on a comparison of
published data with new data obtained on the 6-m telescope. In subsequent
years, several authors studied this variability using high resolution
spectroscopy. Oudmaijer and Bakker [\cite{Bakker}] used a very large
collection of spectrograms with high time resolution and high S/N ratio,
and concluded that there was $\rm V_r$ variability on timescales of
several months in the interval $\rm V_r$\,=\,(84$\div$87)$\pm$2\,km/s,
with an absence of variations with characterisctic timescales from minutes
to hours.

L'ebre et al. [\cite{Lebre}] carried out detailed spectral monitoring of
HD\,56126. Applying a Fourier analysis to radial-velocity data together
with data for the brightness variability, they concluded that the
dynamical state of the atmosphere of HD\,56126 and the pattern inherent to
pulsating RV\,Tau variables were similar. They interpreted the H$\alpha$
variability as a result of shocks. Later, L'ebre et al. [\cite{Lebre2001}]
studied the variability of the H$\alpha$ and H$\beta$ lines. Having
augmented the spectral data and used also photometric observations, they
determined the period of radial pulsations to be P\,=\,36$\lefteqn{.}^{\rm
d}$8. Barth'es et al. [\cite{Barthes}] collected all reliable $\rm V_r$
measurements for HD\,56126 (89 values obtained over eight years) and
analyzed them with photometric data, revealing the presence of $\rm V_r$
variability with a half-amplitude of 2.7\,km/s and the primary period
P\,=\,36.8$\pm 0\lefteqn{.}^{\rm d}2$. The photometric variability
displays the same period, with a very low amplitude (0$\lefteqn{.}^m$02).
They concluded that the star's variability differs substantially from the
pulsations observed in RV\,Tau stars. With its fairly high temperature,
HD\,56126 has evolved further than the RV\,Tau stage. The variability of the
brightness and radial velocity of HD\,56126 may be due to first-overtone
radial pulsations caused by shocks that generate complex, asynchronous
motions in the upper hydrogen-rich layers of the star.

Our new $\rm V_r$ data for HD\,56126 are summarized in the Table. Taking
into account the large probability of differential shifts in the outer
layers of the stellar atmosphere, we present here $\rm V_r$ values for
individual lines and groups of lines. As we can see, the velocity
variations are real, even when they are detected for weak absorption lines
(R$_\lambda \rightarrow $0). The positions of circumstellar features (the
NaI D line and C$_2$ Swan band; see columns~8 and 9 of the Table) display
good consistency. Their comparison with the data from column (4) yields an
expansion velocity for the envelope $\rm V_{exp} \approx$\,11\,km/s.
Note that the position of the primary H$\alpha$ component is not always
consistent with the positions of the D NaI lines and Swan band, whereas
the behavior of the H$\beta$ line is synchronized with the FeII(42) line.
There are still no sufficient reasons to consider HD\,56126 a binary;
however, this cannot be conclusively ruled out, either. In this
connection, it is worth noting the remark made by Barth'es et al. [\cite{Barthes}]
concerning the weak trend of the star's radial velocity during their
long-term observations. This trend may provide evidence for the presence
of a second component in the system, with the orbital period exceeding 16
years. Our eight spectra, obtained at later dates, cannot clarify this
situation. The variability of the velocities derived from extremely weak
absorption lines, $\rm V_r$(R$_\lambda \rightarrow $0), may indicate binarity of the star, but
it may also be a manifestation of low-amplitude pulsations in
circum-photospheric layers. Resolving the question of binarity for
HD\,56126 requires that the behavior of $\rm V_r$ be followed over several
years, with one or two spectra obtained regularly each month.

\subsection{Asymmetry of the profiles and differential shifts of the lines}

When comparing our $\rm V_r$ values with those from other studies, we
should bear in mind not only the different methods used, but also the
spectroscopic peculiarity of the object, and the asymmetries of lines in
the spectrum of HD\,56126 (Figs.~5 and 6). The line profile shapes vary
with both time and line intensity. Therefore, we will consider separately
the data for the line cores in columns~4--8 of the Table (i.e., the lower
parts of the absorption lines or the absorption components of the
H$\alpha$ profiles; or sometimes, in the case of line splitting, individual
components), and supplement these with the $\rm V_r$ estimates derived from
the upper parts and edges of the profiles.

The large spectral interval recorded with the echelle spectrograph made it
possible to study the differential radial-velocity patterns measured from
lines with different intensities. For each of our spectra, we constructed
the dependence $\rm V_r$(R) of the heliocentric radial velocity $\rm V_r$
measured in the absorption core on its central depth R. Figure~7 shows
some of these dependences characteristic of HD\,56126. The velocity of the
center of mass of the star $\rm V_{sys}$\,=\,86.1\,km/s [\cite{Bujar}]
obtained from CO observations at millimeter wavelengths is marked by the
horizontal dashed lines.

Most of the points in these graphs are confined within narrow, almost
horizontal bands, in which the vertical scatter is determined by the
uncertainties for individual lines. The right ends of the bands,
corresponding to the strongest lines formed in the outer layers of the
atmosphere, are sometimes turned upwards or downwards. For $\alpha$\,Per,
the $\rm V_r$(R) dependences are strictly horizontal over the total
interval of line depths. When the C$_2$ molecular lines of HD\,56126
appear in the studied spectral interval, the corresponding points also
form short strips below the main bands, always at the same $\rm V_r$ level
of about 77\,km/s, as was noted above. In contrast, points corresponding
to lines of atoms and ions can occur at, above, or below the level of $\rm
V_{sys}$ (Fig.\,7). Column~4 of the Table contains the average $\rm V_r$
values for the left edges of the main bands; i.e., the limiting values to
which they tend when the central absorption intensities approach the
continuum level (their depths R$_\lambda \rightarrow $0).

Obviously, these can be considered the deepest layers of the atmosphere
accessible to observations. Radial velocities derived from
moderate-intensity absorption lines in the spectrum of HD\,56126 are
presented by L'ebre et al. [\cite{Lebre}] and Barth'es et al.
[\cite{Barthes}]. These values were obtained from only two lines,
BaII\,5854 and CI\,6588\,\AA{}, but for several dozen observing epochs;
the total range of variation of $\rm V_r$ is 81$\div$93\,km/s. Both of
these lines lie on almost horizontal sections of our $\rm V_r$(R)
dependences, to the left of the sharp breaks at R$\approx$60 (Fig.\,7c).
The $\rm V_r$ values measured from our eight spectra that contain at least
one of these two lines vary from 84 to 91\,km/s, within the interval
obtained by Barth'es et al. [\cite{Barthes}]. As we can see from Column~4
of the Table, the limits for the variations of the average velocities
derived from the weakest absorption lines are somewhat lower: 82.5\,km/s$<
\rm V_r$(R$_\lambda \rightarrow $ 0)$<$89.8\,km/s ($\pm$4\,km/s relative to
$\rm V_{sys}$). The average values of our $\rm V_r$(R$_\lambda \rightarrow $ 0)
and for the two lines analyzed in [\cite{Barthes}] coincide within the
errors (87.3$\pm$0.7 and 87.7$\pm$0.3\,km/s). The $\rm V_r$ values
obtained from the stellar components of the NaI D lines (column~8 of the
Table, the value in parentheses) are close to $\rm V_r$(R$_\lambda
\rightarrow $0), but less accurate.

Strong SrII and YII lines and other strong lines (with depths $\rm
R_\lambda \approx 60-70$; points on the $\rm V_r$(R) dependences to the
left of the breaks) display largeamplitude time variations and a
predominance of red shifts relative to the weakest absorption lines:
83\,$<\rm V_r<$\,94 km/s ($-3$\,km/s, +8 km/s relative to $\rm V_{sys}$).
Both of these tendencies are even more pronounced in the FeII(42) and
H$\alpha$ lines (columns~5 and 6 of the Table): 77\,$<\rm V_r<$\,98\,km/s
($-9$\,km/s and +12\,km/s relative to $\rm V_{sys}$). In three cases, the
velocity could also be estimated from reliably detected blue-shifted
components of the FeII(42) lines. The H$\alpha$ profile displays the
highest variability, but for any profile presented in Fig.\,3, an
analogous profile can be found among those presented by L'ebre et al.
[\cite{Lebre}] and Barth'es et al. [\cite{Barthes}], in terms of both
profile shape and the velocities derived from the absorption components.
In our spectra, $\rm V_r$(H$\alpha$) (column~7 of the Table) varies from
54 to 88\,km/s ($-32$ and +2\,km/s, relative to $\rm V_{sys}$) for the
main (deeper) components and from 43 to 112\,km/s ($-43$ and +26\,km/s
relative to $\rm V_{sys}$) for the secondary components.

All the lines except those of hydrogen have the same type of asymmetry:
their blue wing is extended more than the red, i.e., the velocity measured
from the upper part of the profile (from the wings) is smaller than the
velocity measured from the core. The blue shift of the wings relative to
the core increases with line strength, either gradually or jump-like. In
three of our spectra, obtained on November 20, 2002, March 9, 2004, and
November 12, 2005, the absorption lines remain symmetrical up to
R$_\lambda \approx 50$ (horizontal sections of $\rm V_r$(R) dependences in
Fig.\,7), but also display the above asymmetry for the deepest lines
(Figs.\,5 and 6). The FeII(42) absorption lines are asymmetrical for all
our observing epochs (Fig.\,6); on November 14, 2003 and November 12, 2005,
the shift of their wings relative to their cores reaches $-11$\,km/s. In
contrast, in the spectrum obtained on March 4, 1999, the asymmetry is seen
even in the weakest lines; in the interval 5\,$< \rm R <$\,50, as the
lines deepen and their blue shifts grow, the difference between the $\rm
V_r$ values measured from the wings and cores increases from $-3$ to
$-6$\,km/s, but the most blue-shifted FeII(42) absorption lines are again
almost symmetrical (Figs.\,6 and 7a). The wings of strong lines are less
shifted relative to the weak lines than are their cores, and the velocities
derived from the wings are closer to the center-of-mass velocity of the
star (deviations from $\rm V_{sys}$ are from $-10$ to +3\,km/s). We can
see from Fig.\,3 and the figures presented in [\cite{Barthes}] that both
the blue the red slopes of the H$\alpha$ absorption component may be
flatter (we are speaking here of the central part of the profile, rather
than the broad photospheric wings).

Strong absorption lines in the spectrum of HD\,56126 display fairly
clear-cut boundaries, from which the velocities were also measured. These
measurements show that the red boundary of the profile is substantially
more stable than the blue. For the red boundaries of the strongest
non-hydrogen lines, it is sufficient to indicate the average velocity,
namely $\rm V_r$\,=\,124$\pm$2\,km/s, while the velocity for the left
boundaries ranges from 25 to 53\,km/s. For H$\alpha$, the same estimates
yield $\rm V_r$\,=\,133$\pm$5\,km/s and 10$\div$55\,km/s. Thus, the
deviations from $\rm V_{sys}$ are $-33\div -61$\,km/s and +38$\pm$2\,km/s
in the former and $-31\div -76$\,km/s and +47$\pm$5\,km/s in the latter
case.

Summarising the measurements and analysis of the radial velocity
pattern, we draw the following conclusions.

\begin{itemize}

\item{Our data for the radial velocities of HD\,56126 are fairly accurate
(the systematic errors are below 1\,km/s), making it possible to combine
them with the most accurate data obtained previously.}

\item{We have found substantial differential shifts for lines of different
intensities within the same spectrum (up to 15\,km/s for metallic lines,
and up to 30\,km/s for H$\alpha$); these are the $\rm V_r$(R) dependences
manifesting the $\rm V_r$ gradient. The time variability of the $\rm V_r$(R)
dependences indicates that velocities should be measured using a large set
of lines within an extended spectral interval.}

\item{Both expanding matter and matter falling onto the star are seen in
      the stellar atmosphere.}

\item{Data for extremely weak absorption lines are important ($\rm
V_r(R\rightarrow$0)) for studies of the velocity pattern; the detected
variability of these velocities, 82.5$<\rm V_r (R\rightarrow 0)
<$\,89.8\,km/s, may either represent low-amplitude pulsations in the
circumphotospheric layers or indicate the star's binarity.}

\item{We have found complex and variable profiles of strongest lines (not
only hydrogen, but also FeII, YII, BaII, and other absorption lines)
formed in the expanding stellar atmosphere (at the base of the wind).
Measurements for the velocities derived from individual features of these
profiles are needed for kinematic studies.}

\item{We have confirmed the stability of the expansion velocity for the
circumstellar envelope of HD\,56126, determined from the C$_2$ and D2 lines
of NaI.}

\item{High and ultra-high spectral resolution is needed to study,
respectively, stellar and circumstellar lines in the spectrum of HD\,56126
      displaying complex structures.}

\end{itemize}

\section{Conclusions}

Our high resolution spectra of HD\,56126 have revealed variability of
various line profiles: along with the previously known H$\alpha$ profile
variability, the profiles of strongest lines (such as BaII, YII, and FeII)
also proved to be variable.
The broad spectral range encompassed by our spectra enabled us to measure
radial velocities using spectral features that form at various depths in
the stellar atmosphere and circumstellar envelope.

We were able to distinguish the C$_2$ absorption molecular bands
(and their structure), as well as bands identified with diffuse
interstellar bands. We found substantial differential shifts in lines of
different intensities within the same spectrum, i.e., appreciable $\rm
V_r$(R) dependences, as well as variation of these shifts with time. This
indicates the need for velocity measurements based on a large set of lines
in an extended spectral interval.

We conclude that both expanding layers and matter falling onto the star
exist simultaneously in the stellar atmosphere. The position of the
molecular spectrum is stable, indicating stability of the expansion
velocity of the envelope.

\section*{Acknowledgements}

This work was supported by the Russian Foundation for Basic Research
(project code 05--07--90087), the program of basic research held by the
Presidium of Russian Academy of Sciences ``The origin and evolution of
stars and galaxies'', program held by the Section for Physical sciences of
RAS ``Extended objects in the Universe''. This publication is based on
work supported by Award No.RUP1--2687--NA--05 of the U.S. Civilian
Research \& Development Foundation (CRDF). In our study we used SIMBAD and
ADS databases.

\newpage

\begin{figure}[t]	      		      
\includegraphics[angle=0,width=1.0\textwidth,bb=0 0 595 460,clip]{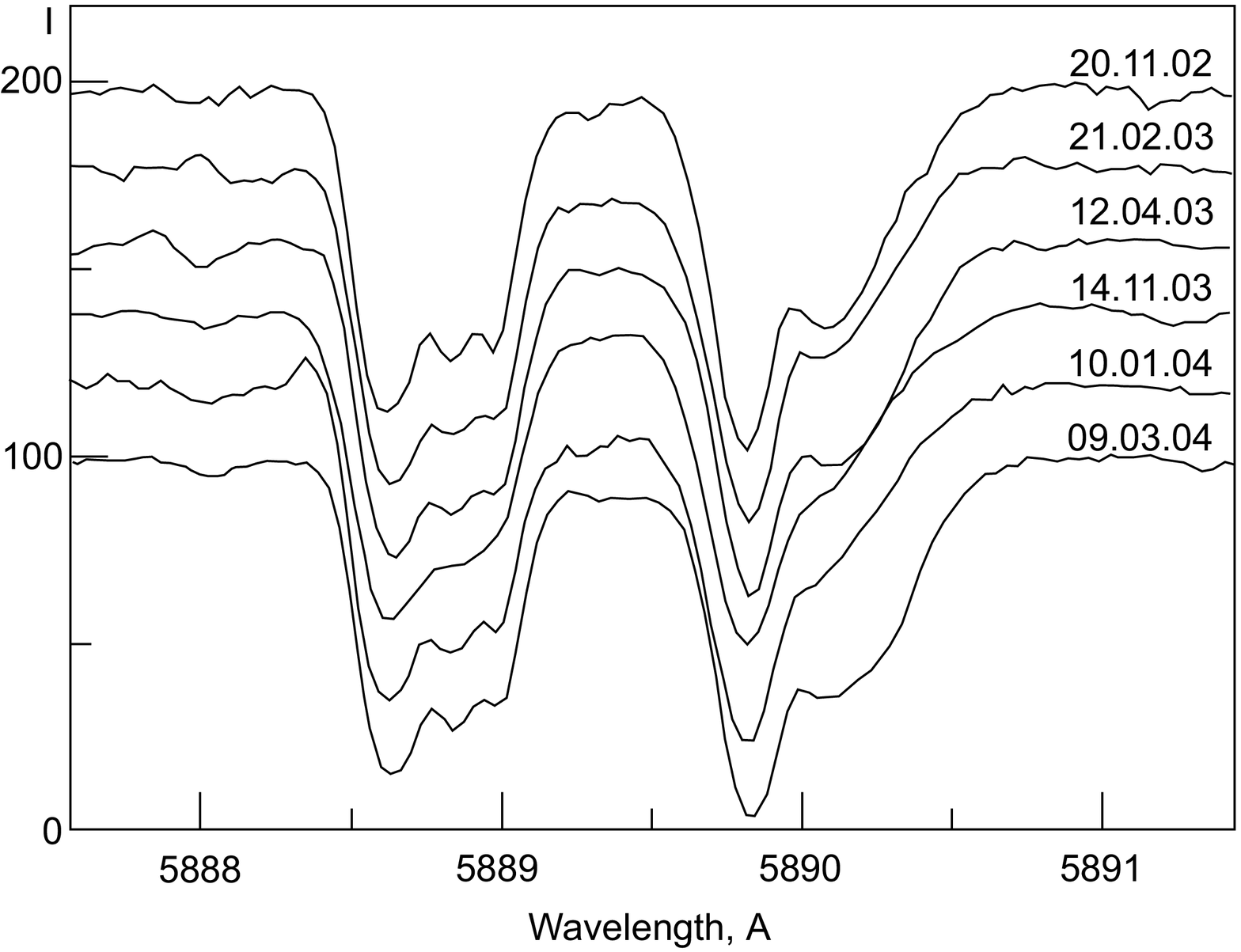}
\caption{Spectral fragments containing the D2 NaI line for various observing dates}
\end{figure}

\newpage
\begin{figure}[t]	      		      
\includegraphics[angle=0,width=1.0\textwidth,bb=10 10 310 210,clip]{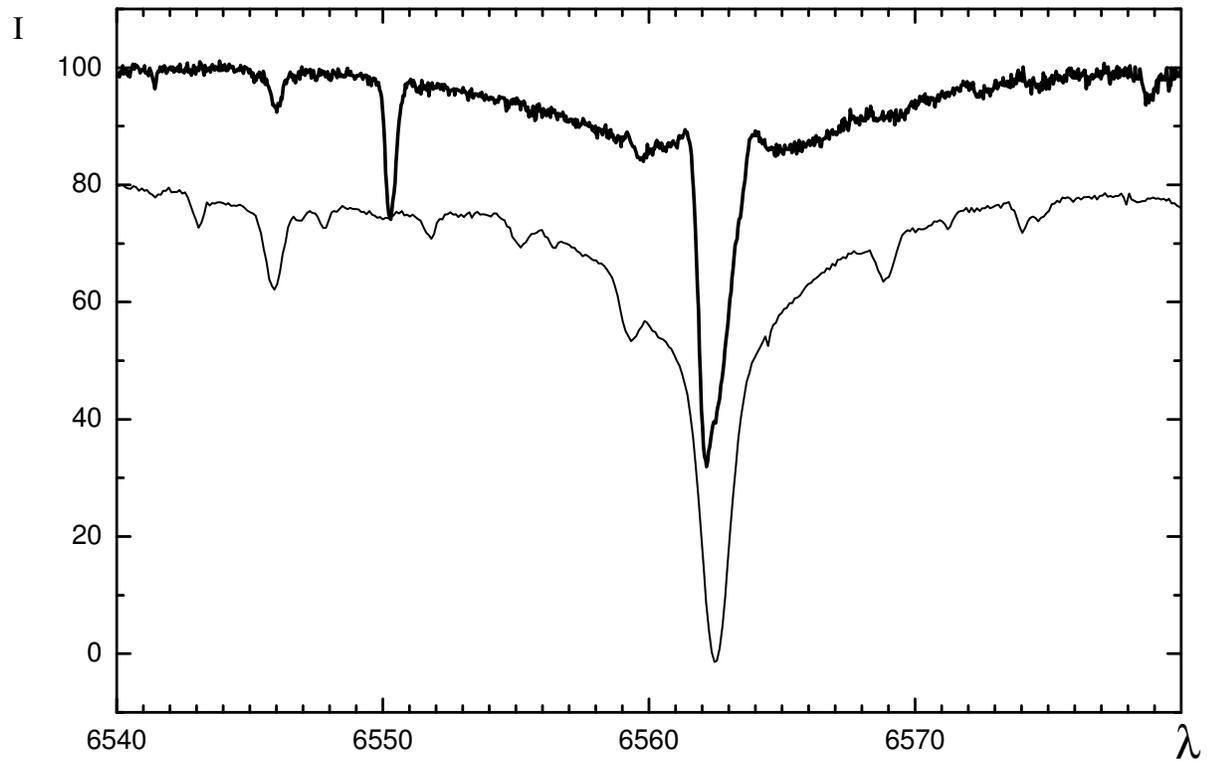}
\caption{H$\alpha$ profile in the spectra of HD\,56126 (top) and $\alpha$\,Per
           (bottom).}
\end{figure}

\begin{figure}[t]	      		      
\includegraphics[angle=0,width=0.5\textwidth,bb=1 1 360 730,clip]{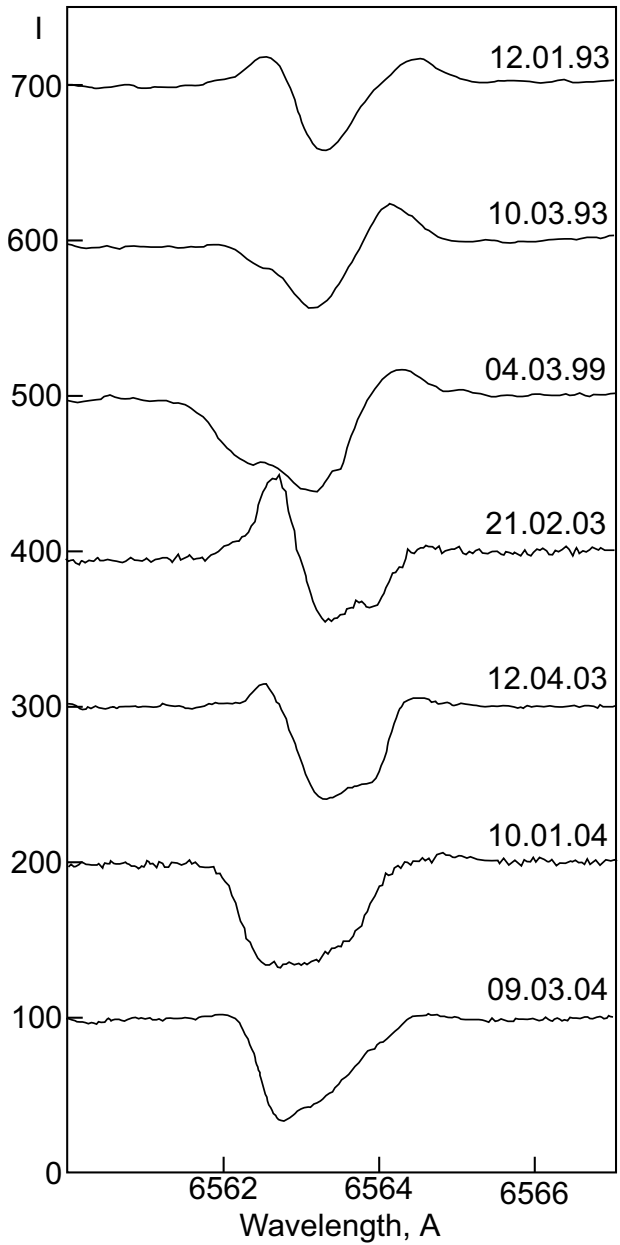}
\caption{Central part of the H$\alpha$ profile in the spectra of HD\,56126
        obtained on different nights.}
\end{figure}

\begin{figure}[t]	      		      
\includegraphics[angle=0,width=1.0\textwidth,bb=1 1 310 210,clip]{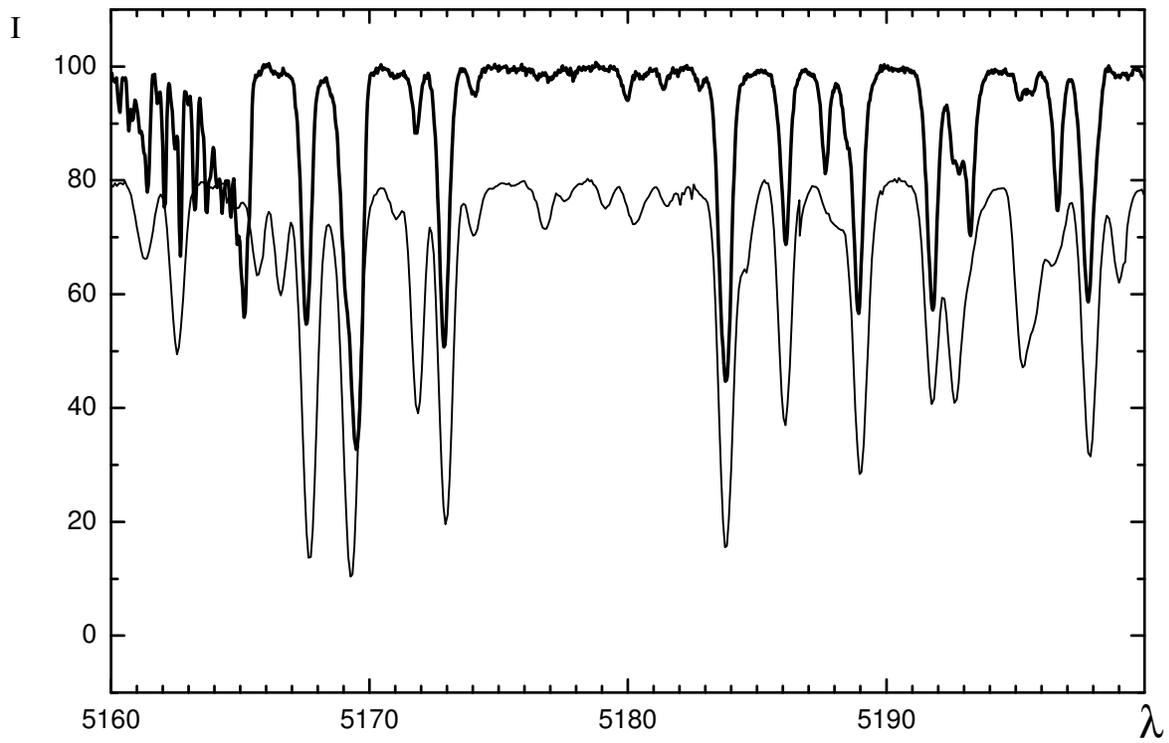}
\caption{Fragment of the spectrum of HD\,56126 with the Swan band 5165\,\AA{} 
        of the C$_2$ molecule. For comparison, the same spectral fragment for
	$\alpha$\,Per is presented below.}
\end{figure}

\begin{figure}[t]	      		      
\includegraphics[angle=0,width=0.6\textwidth,bb=1 1 490 720,clip]{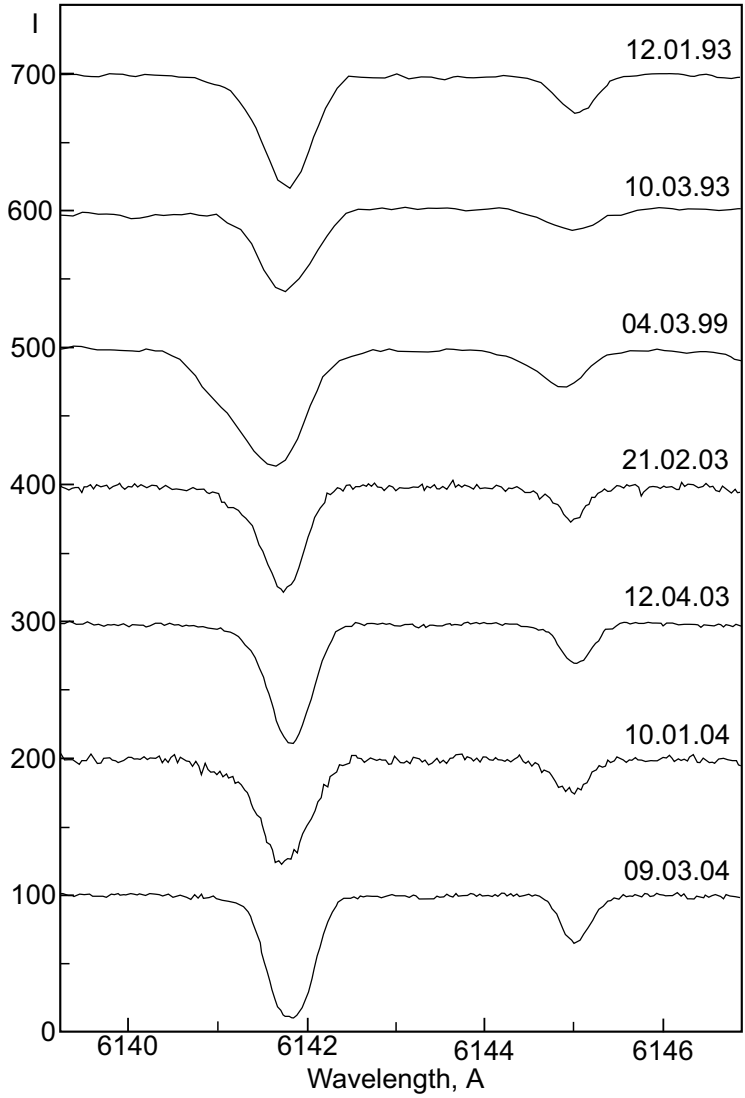}
\caption{Variability of the BaII 6141\,\AA{} line profile in the spectrum of HD\,56126.}
\end{figure}

\begin{figure}[t]	      		      
\includegraphics[angle=0,width=0.7\textwidth,bb=1 1 540 640,clip]{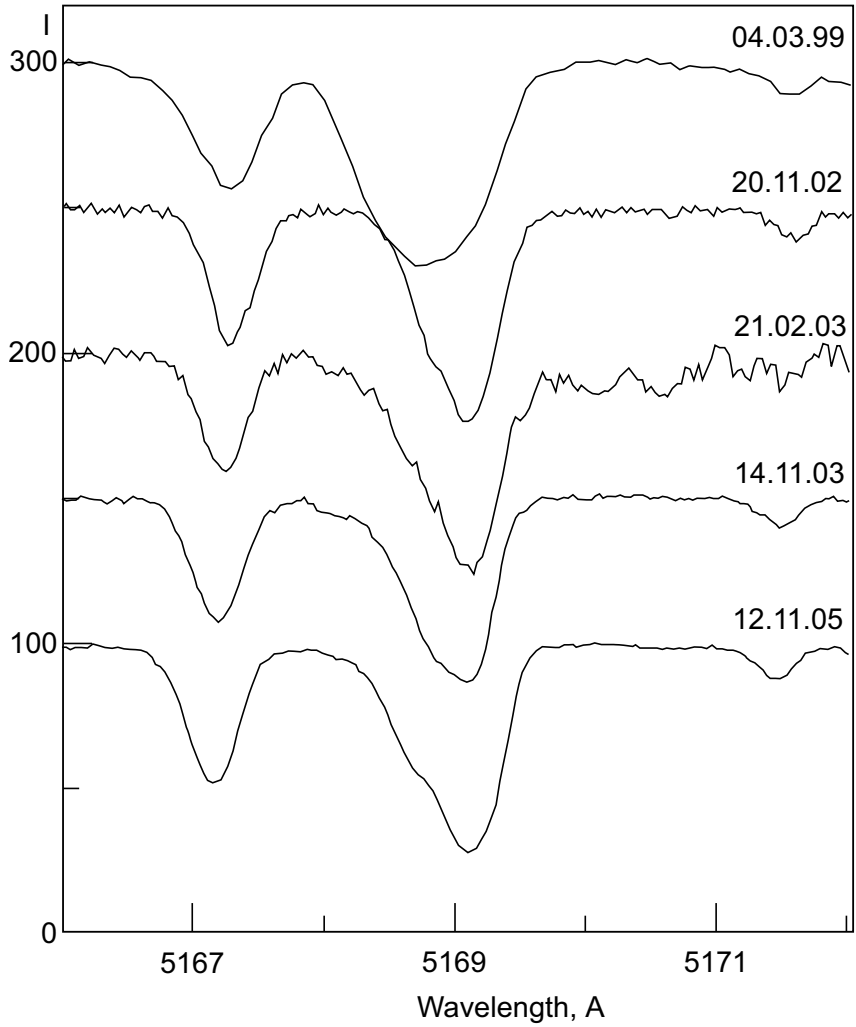}
\caption{Variability of the FeII 5169\,\AA{} line profile in the spectrum
         of HD\,56126.}
\end{figure}

\voffset=-2cm
\begin{figure}[t]	      		      
\includegraphics[angle=-90,width=0.5\textwidth,bb=50 130 550 800,clip]{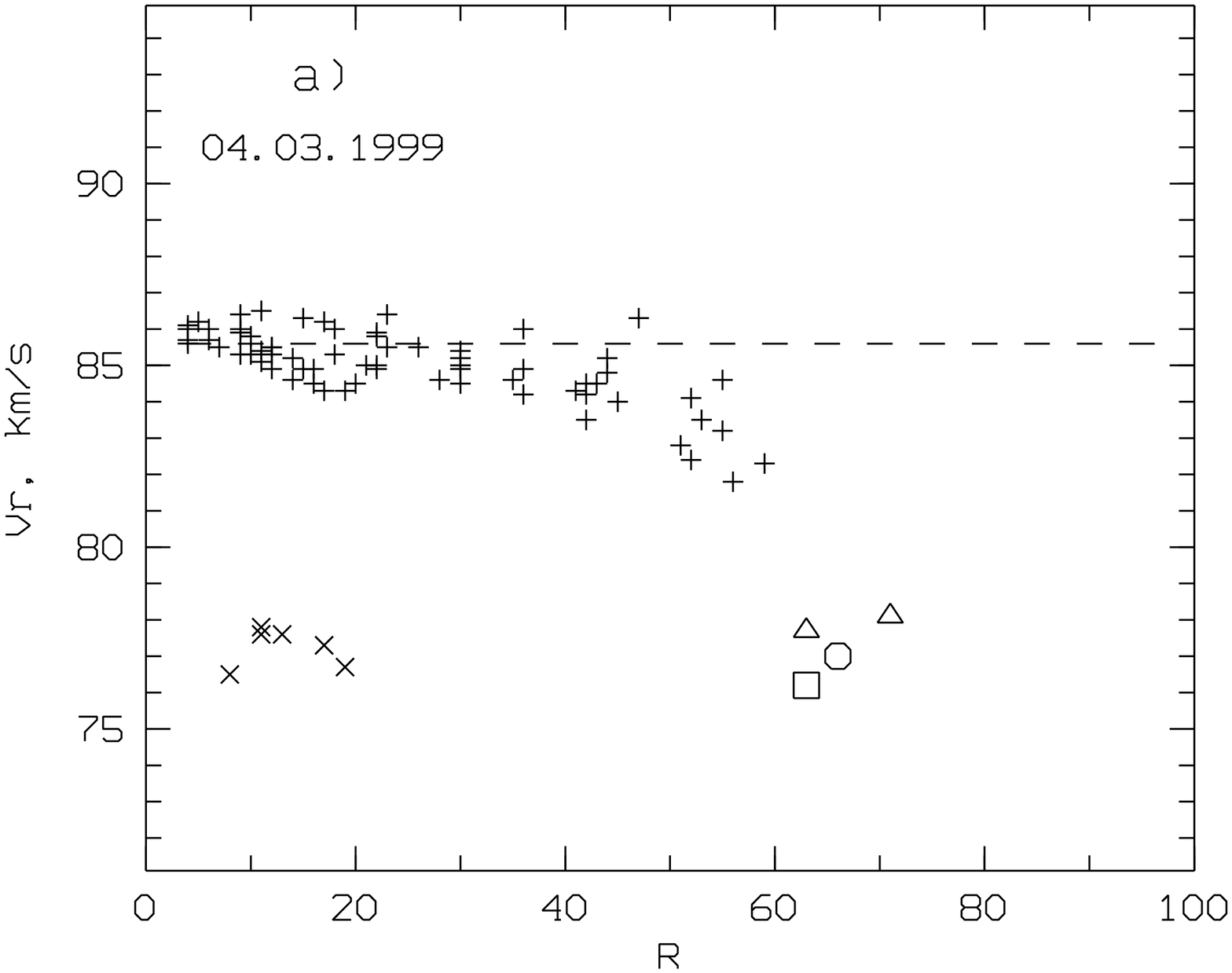}

\includegraphics[angle=-90,width=0.5\textwidth,bb=50 130 550 800,clip]{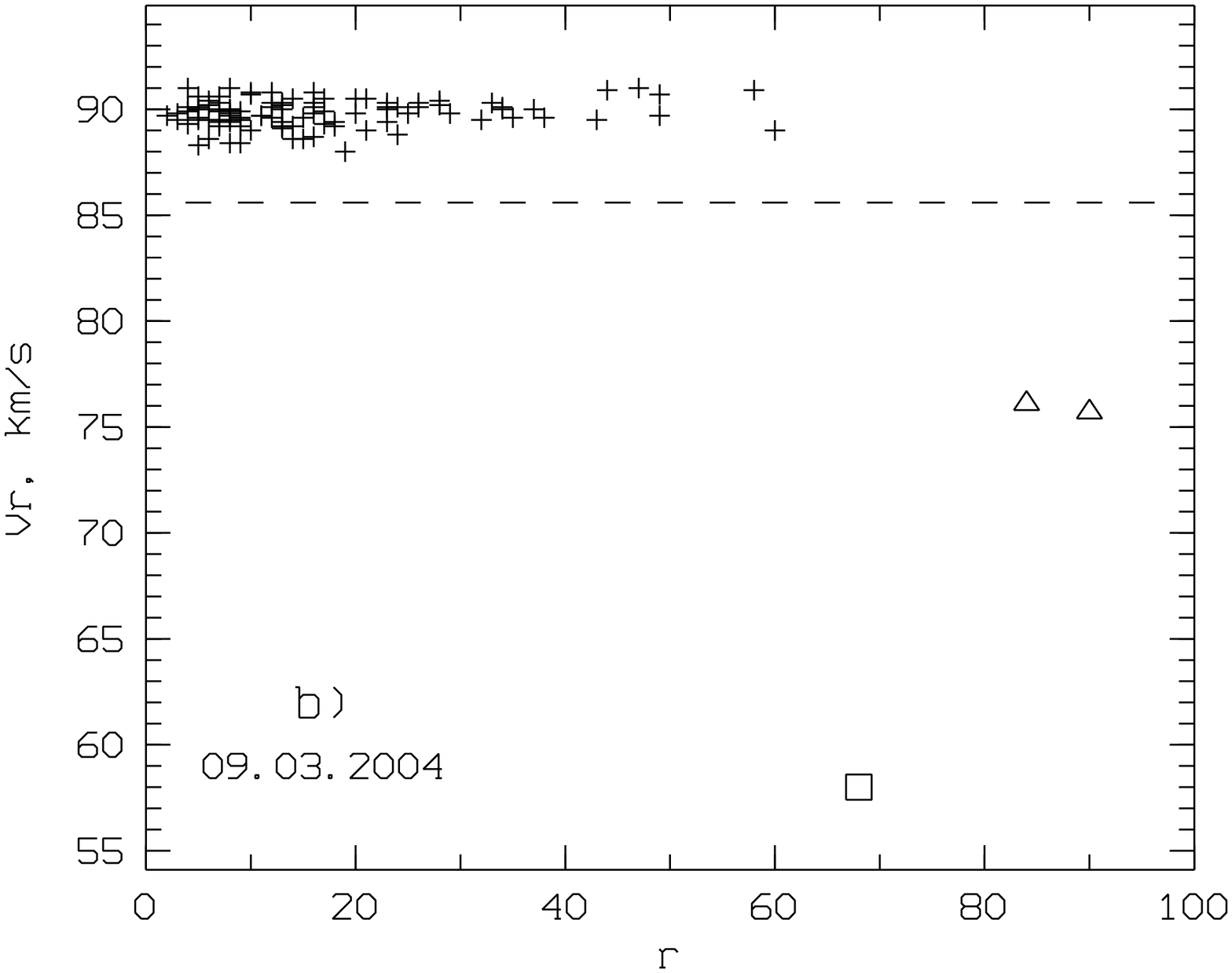}

\includegraphics[angle=-90,width=0.5\textwidth,bb=50 130 580 800,clip]{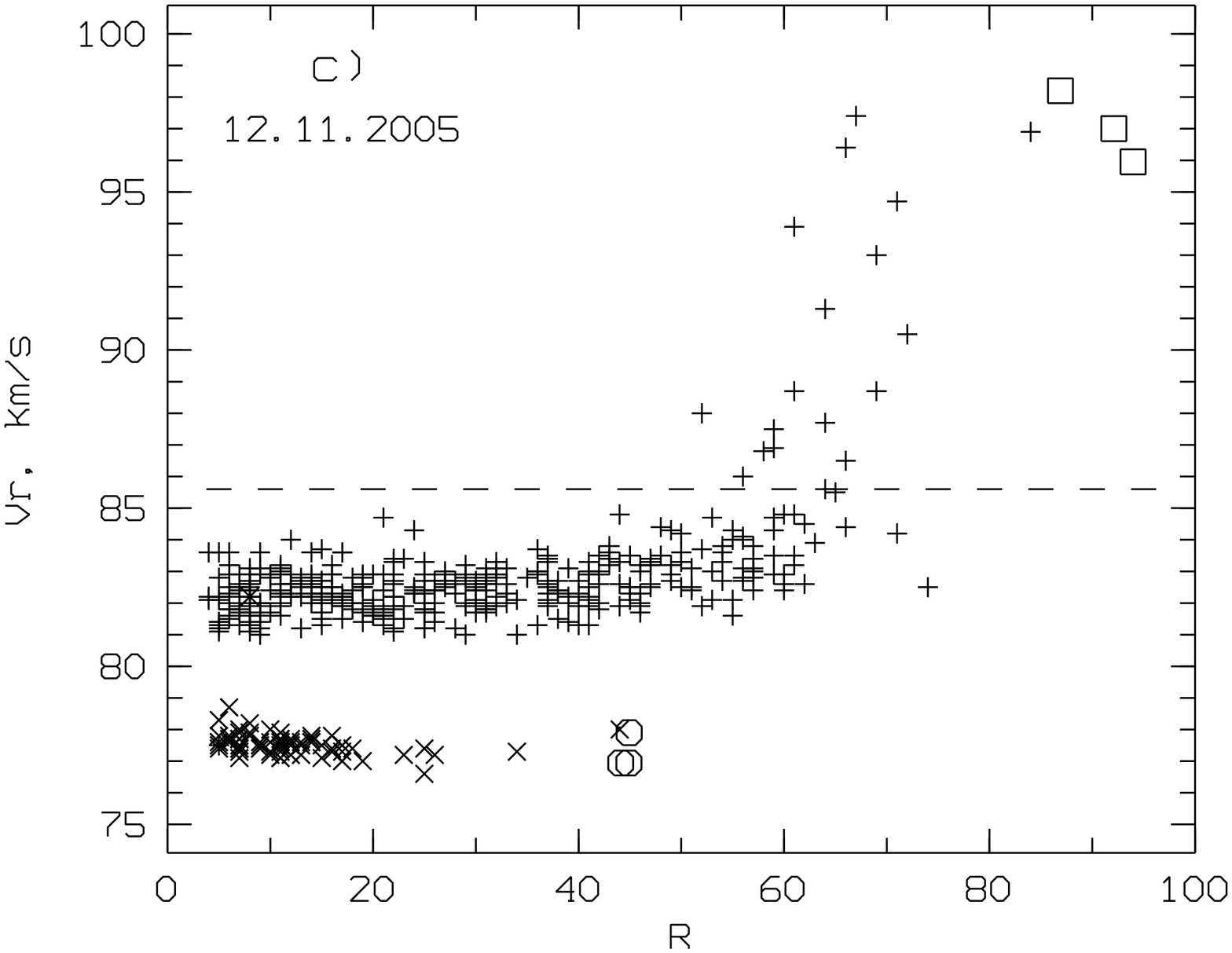}
\caption{Dependence of the heliocentric radial velocity derived from the absorption
         core on its depth R in the spectrum of HD\,56126 for various nights:
	 (a) March 4, 1999, (b) March 9, 2004, (c) November 12, 2005.
	 The dashed line indicates the systemic velocity. Each symbol denotes
	 a particular type of spectral line: the pluses denote lines of metals;
	 the daggers --  C$_2$ lines; and the triangles, circle, and square --
	 the main components of the D2 NaI, FeII(42) 5169\,\AA{}, and H$\alpha$ lines,
	 respectively. In graph (c), the circles denote the secondary components
	 of the FeII(42) lines, and the squares the H$\beta$, H$\gamma$, and H$\delta$
	 lines (top to bottom). The size of the symbols corresponds to the accuracy
	 of the $\rm V_r$ measurements and the depth of the lines.}
\end{figure}

\end{document}